\documentclass[12pt]{article}
\usepackage{amsmath,bm}
\usepackage{setspace}
\usepackage{verbatim}
\usepackage{fullpage}
\usepackage{amsthm,amsfonts,amssymb,bm,epsfig,psfrag}
\newcommand{\im}{\textrm{Im}}
\newcommand{\re}{\textrm{Re}}

\newcommand{\pif}{P_{i\rightarrow f}}
\newcommand{\Hessp}{\nabla^{2} \pif}
\newcommand{\pp}{P_{1\rightarrow 5}}
\newcommand{\trans}{\mathsf{T}}

\newcommand{\control}{\beta}
\newcommand{\Rn}{\mathbb{R}^{M}}

\newcommand{\level}{\mathcal{L}}

\newcommand{\Pcrit}{\hat{P}^{\textrm{c}}}

\newcommand{\gradp}{\nabla \pif}

\newcommand{\gradJ}{\nabla J}
\newcommand{\betastar}{\beta^{*}}
\begin{document}
\doublespace
\center{\Large{Exploring the top and bottom of the quantum control landscape}}\\
Vincent Beltrani,$^{a)}$ Jason Dominy,$^{b)}$ Tak-San Ho,$^{a)}$ and Herschel
Rabitz$^{a)}$ \\
\begin{center}
$^{a)}$ \small{Department of Chemistry, Princeton University, Princeton NJ 08544
}\\
$^{b)}$ \small{Program in Applied and Computational Mathematics, Princeton
University, Princeton NJ 08544}\\
\small{(Received} 
\end{center}

\abstract{A controlled quantum system possesses a search landscape defined by
the target physical objective
as a function of the controls.  This paper focuses on the landscape for the
transition probability $\pif$ between the states of a finite level quantum system. 
Traditionally, the controls are applied fields; here we extend
the notion of control to also include the Hamiltonian structure, in the form of
time independent matrix elements.  Level sets of controls that produce the same
transition probability value are shown to exist at the bottom $\pif=0.0$ and top
$\pif=1.0$ of the landscape with the field and/or Hamiltonian structure as controls.  
We present an algorithm to continuously explore
these level sets starting from an initial point residing at either extreme value
of $\pif$.  The technique can also identify control solutions that exhibit the
desirable properties of (a) robustness at the top and (b) the ability to rapidly
rise towards an optimal control from the bottom.  Numerical simulations are
presented to illustrate the varied control behavior at the top and bottom of the
 landscape for several simple model systems.  
}
\clearpage

\section{Introduction}
Control of quantum phenomena is garnering increasing attention as growing numbers
of experimental studies
indicate the ability to find effective control over a broad range
of quantum 
systems \cite{rabitz:quant:opt}.  Successful experiments using shaped laser
pulses include selective 
bond breaking \cite{assion}, selective excitation of extremely similar species
\cite{roth:253001}, control of optical switches in semiconductors \cite{kunde}, 
selective energy transfer in bio-molecules \cite{herek}, the manipulation
of electron excitation \cite{weinacht}, etc. 

The control landscape is defined as the physical observable as a function of the control variables.
The general structure of the landsacpe dictates the ease of finding good controls as well as robustness to noise
and other characteristics of a controlled quantum system.  Thus, 
exploration of the landscape is important, and the tools developed in this work are applicable to general
quantum observables.  For illustration, this paper considers the transition
probability, $\pif$, from state $|i\rangle$ to $|f\rangle$.  The transition
probability landscape has been shown 
\cite{kate:pif,shen,rothman:dmorph,rothman:level,ash} to generally
exhibit a fundamental trap-free topology for completely controllable quantum
systems, i.e., satisfaction of certain physical assumptions assures that there
are no sub-optimal extrema capable of halting gradient based optimizations
\cite{Jurdjevic1972313,PhysRevA.51.960,981126}.  Large, even infinite, numbers
of control solutions are capable of achieving the same observable value defining
members of a quantum control level set.  Traditionally, the controls have taken
the form of electromagnetic
fields applied to a fixed physical system.  In contrast, a recent study treated
the Hamiltonian structure (i.e., time independent matrix elements) as the
controls \cite{ash}. In this circumstance, the Hamiltonian structure in the laboratory can
be varied,
for example, by substituting one functional group for another bonded to a
molecular scaffold or by altering
the relative composition of a material sample.  Figure 1 shows the selection
options for control variables.  Depending on the control scenario, the applied
field (Fig. 1a) or the Hamiltonian structure (Fig. 1b)  may be viewed as the
control.  The most general scenario (Fig. 1c) considers both the Hamiltonian
structure and the applied field as dual controls.  Exploiting this rich
flexibility could be important in order to meet particularly demanding
objectives  (e.g., prescribed complex responses from a non-linear optical
material).  

Control level sets, over the interior domain $0<\pif<1$, have been investigated
through a first-order (i.e., through the use of the gradient of $\pif$ with
respect to the controls) numerical algorithm called D-MORPH
\cite{rothman:dmorph,rothman:level,rothman:track}.  The solution to the D-MORPH
differential equation permits exploration of the connected members of the
control level sets.  The landscape top $\pif=1.0$ and bottom $\pif=0.0$ have
particular physical relevance.  The top is the most desired location while at
the bottom the goal is to climb away as rapidly as possible.  In order to
explore level sets at the top or bottom, this work generalizes the original first-order
D-MORPH algorithm to produce a new second-order formulation.  This extended
D-MORPH algorithm utilizes the Hessian as the second derivative of $\pif$ with
respect to the control variables.  At both landscape extremes, the Hessian has a
large number of zero eigenvalues and associated eigenvectors that span the
Hessian null space.  The eigenvectors spanning the null space describe local
coordinated variations of the controls that leave the $\pif$ value fixed.  The
second-order D-MORPH algorithm systematically makes such changes of the controls
in these null spaces to continuously explore the level sets at the landscape top
or bottom.  
At this early stage of landscape feature exploration, the myopic nature of
D-MORPH based simulations are important to link with the associated theoretical
analyses and provide a foundation for subsequent laboratory landscape excursions
with a variety of possible algorithms.

The remainder of the paper is organized as follows:
Section 2 presents the general formulation of the gradient and Hessian of the
transition probability with respect to the controls.  Utilizing the gradient and
Hessian, Section 3 summarizes the algorithm used to explore the top and bottom
of the $\pif$ landscape.  Section 4 presents numerical results using simple few
level systems for illustration, and Section 5 offers concluding remarks on the
significance of the work.

\section{Transition Probability Landscape}\label{sec:pif}
Throughout the paper, the symbol $\control$ will be used to indicate
collectively the set of all the control parameters in the system $\beta_{j},
j=1,2,\dots,M$, i.e., $\control\in\Rn$.  These $M$ control parameters may be
drawn from any component of the Hamiltonian under consideration. For example,
these parameters may consist of the field spectral amplitudes and phases, as
well as the Hamiltonian structure itself, i.e., the dipole matrix elements and
system energy levels.  The collective controls, $\beta$, can be considered
within the same unified framework regardless of their specific nature in the
Hamiltonian which will be written as $H(\control,t)$. 
In the remainder of the paper, a control that produces an interior yield $0 <
\pif < 1$ will be denoted as $\beta$, while a  control producing either $\pif =
0.0$ or $\pif = 1.0$ will be written as $\beta^{*}$. 

The transition probability landscape is defined by $\pif(\beta)$ as a function
of $\beta$
\begin{equation}\label{pif_eq}
\pif(\control) = |\langle f| U(T,0)| i\rangle|^{2}.
\end{equation}
Here, $U(T,0)$ is the propagator evaluated at the target time $T$, governed by
the Schr\"{o}dinger equation
\begin{equation}\label{schrod_nom}
i\hbar \frac{\partial}{\partial t} U(t,0) = H(\control,t)U(t,0), \quad U(0,0)=1.
\end{equation}
The propagator $U(t,0)$ is implicitly dependent on $\beta$ through the
Hamiltonian $H(\beta,t)$ in Eq. \ref{schrod_nom}.  The system has $N$ states and
we assume that the initial and target states, $|i\rangle$ and $|f\rangle$,
respectively, are orthogonal, $\langle f|i\rangle = 0$.

Below, we construct the first and second derivatives of the propagator $U(t,0)$
with respect to the parameters $\beta$, as they are needed for the remainder of
the landscape analysis.  On differentiating Eq. \ref{schrod_nom} with respect to
an arbitrary parameter $\beta_{j}$, we obtain
\begin{equation}\label{inhomo}
 i\hbar \frac{\partial}{\partial t} U_{\beta_{j}}(t,0) =
H_{\beta_{j}}(\beta,t)U(t,0) + H(\beta,t)U_{\beta_{j}}(t,0),
\end{equation}
with initial conditions $U_{\beta_{j}}(0,0) = 0$, where we use the notation
$U_{\beta_{j}}(t,0)\equiv \partial U(t,0)/\partial\beta_{j}$ and
$H_{\beta_{j}}(\beta,t)\equiv \partial H(\beta,t)/\partial \beta_{j}$, .  Equation
\ref{inhomo} is an inhomogeneous equation with a homogeneous part identical to
Eq. \ref{schrod_nom}.  Thus, the solution to Eq. \ref{inhomo} can be written as
\cite{green, taksan:control:easy}
\begin{equation}\label{Uderiv}
 U_{\beta_{j}}(t,0) = (-i/\hbar) U(t,0) \int_{0}^{t}  \ dt' \ U^{\dag}(t',0)
H_{\beta_{j}}(\beta,t') U(t',0).
\end{equation}

The second order derivatives $U_{\beta_{i}\beta_{j}}(t,0) \equiv
\partial^{2}U(t,0)/\partial\beta_{i}\partial\beta_{j}$ can be derived similarly
by differentiating Eq. \ref{inhomo} and utilizing Eqs. \ref{schrod_nom} and
\ref{Uderiv}.  Here, we only present the result:
\begin{align}\label{Usecond}
 U_{\beta_{i}\beta_{j}}(t,0) &= (-i/\hbar) U(t,0) \int_{0}^{t} \ dt' \
U^{\dag}(t',0)
[ H_{\beta_{i}\beta_{j}}(\beta,t')U(t',0) \nonumber \\
&\quad\quad + H_{\beta_{i}}(\beta,t')U_{\beta_{j}}(t',0) + 
H_{\beta_{j}}(\beta,t')U_{\beta_{i}}(t',0)],
\end{align}
where $H_{\beta_{i}\beta_{j}}(\beta,t)\equiv
\partial^{2}H(\beta,t)/\partial\beta_{i}\beta_{j}$.

\subsection{Gradient of the Transition Probability Landscape}
We consider here a control $\beta$ corresponding to a point on the landscape
where $\pif(\beta)$ is not $0.0$ or $1.0$.  
The behavior of the landscape in the neighborhood of such a control $\beta$ can
be explored by expanding $\pif(\beta+d\beta)$ to first order
\begin{subequations}\label{1st_order_exp}
\begin{align}
\pif(\beta+d\beta) &\approx \pif(\beta) + \sum_{j=1}^{M}
\frac{\partial\pif(\beta)}{\partial\beta_{j}} d\beta_{j} \\
 &= \pif(\beta) + \gradp(\beta)^{\trans} \cdot d\beta \label{gradexp}
\end{align}
\end{subequations}
where the first derivative of the transition probability,
$\partial\pif(\beta)/\partial \beta_{j}$, is given by 
\begin{align}
\partial\pif(\beta)/\partial\beta_{j} &= \frac{\partial}{\partial \beta_{j}}
|\langle f|U(T,0)|i\rangle|^{2} \nonumber \\
&= \langle i| U^{\dag}_{\beta_{j}}(T,0)|f\rangle\langle f|U(T,0)|i\rangle + 
\langle i| U^{\dag}(T,0)|f\rangle\langle i| U_{\beta_{j}}(T,0)|i\rangle
\label{almost} \\
&= (2/\hbar) \int_{0}^{T} \im [ \langle \chi|
U^{\dag}(t,0)H_{\beta_{j}}(\beta,t)U(t,0)|i\rangle] \ dt. \label{1st:pif:deriv}
\end{align}
In Eq. \ref{gradexp}, `$\trans$' denotes the vector transpose operation and the
gradient $\gradp(\beta)$ will be used as a compact notation throughout the
paper.  
Equation \ref{Uderiv} was used on going from Eq. \ref{almost} to Eq.
\ref{1st:pif:deriv} where $|\chi\rangle \equiv U^{\dag}(T,0)|f\rangle
\langle f| U(T,0)|i\rangle$.  The first order derivative in Eq.
\ref{1st:pif:deriv} is useful for guiding the control $\beta$ to either extremum
$\pif=0.0$ or $\pif=1.0$.  

\subsection{Hessian of the Transition Probability Landscape}
The top or bottom of the landscape is an extremum with a corresponding control
$\beta^{*}$ satisfying $\partial\pif(\beta^{*})/\partial\beta^{*}_{j} = 0$ for
$j=1,\dots,M$.  The behavior in the neighborhood of an extremum can be explored
by expanding $\pif(\beta^{*}+d\beta)$ to second order 
\begin{equation}\label{expand:2nd}
 \pif(\beta^{*}+d\beta) \approx \pif(\beta^{*}) +
\frac{1}{2}\sum_{j=1}^{M}\sum_{k=1}^{M} d\beta_{j}
\frac{\partial^{2}\pif(\beta^{*})}{\partial\beta^{*}_{j}\partial\beta^{*}_{k}}
d\beta_{k},
\end{equation}
where the second derivatives of the transition probability,
$\partial^{2}\pif(\beta)/\partial\beta_{i}\partial\beta_{j}$, may be calculated
by differentiating Eq. \ref{1st:pif:deriv} with respect to $\beta_{k}$, leading
to the general expression   
\begin{align}\label{second:gen}
 \partial^{2}\pif(\beta)/\partial\beta_{i}\partial\beta_{j} &= 
2\re\{ \langle i| U^{\dag}(T,0)|f\rangle\langle f|
U_{\beta_{i}\beta_{j}}(T,0)|i\rangle \nonumber \\
&\quad\quad + \langle i| U^{\dag}_{\beta_{i}}(T,0)|f\rangle \langle f|
U_{\beta_{j}}(T,0)|i\rangle \}.
\end{align}
Equation \ref{second:gen} applies at any point on the landscape and Sections
\ref{sec:hess:top} and \ref{sec:hess:bott} will, respectively, specialize this result to cases at the top
and bottom.  We can conveniently write Eq. \ref{expand:2nd} in matrix-vector
form as
\begin{equation}\label{eq:sec:exp}
 \pif(\beta^{*}+d\beta) \approx \pif(\beta^{*}) + (1/2) \ d\beta^{\trans} \cdot
\Hessp(\beta^{*}) \cdot d\beta.
\end{equation}
In Eq. \ref{eq:sec:exp}, $\Hessp(\beta^{*})\in\mathbb{R}^{M\times M}$ is the
Hessian matrix, i.e., the matrix of second partial derivatives with respect to
the system parameters, evaluated at $\beta^{*}$.  

\subsubsection{Hessian at the Top of the Landscape}\label{sec:hess:top}
Previous formal work \cite{taksan:control:easy, rabitz:top:pif} showed that
multiple control solutions can exist at the top of the landscape where
$\pif=1.0$.  The Hessian at the top can facilitate exploration of the family of
control solutions there.  In particular, at the top of the landscape,
$U(T)|i\rangle = e^{i\theta}|f\rangle$ for some phase $\theta\in[0,2\pi)$, and
substituting this relationship into Eq. \ref{second:gen} upon using Eqs.
\ref{Uderiv} and \ref{Usecond} leads to the result
\begin{equation}\label{hess:elements}
\begin{split}
\frac{\partial^{2}\pif}{\partial\beta^{*}_{j}\partial\beta^{*}_{k}}
& =(-2/\hbar^{2}) \sum_{p\neq i} \re \Big[ \int_{0}^{T} \langle p|
U^{\dag}(t,0)H_{\beta^{*}_{j}}(\betastar,t)U(t,0)|i\rangle dt   \\
& \quad\quad \times \int_{0}^{T} \langle i| U^{\dag}(t,0) H_{\beta^{*}_{k}}(\betastar,t)
U(t,0)|p\rangle \ dt \Big].
\end{split}
\end{equation}
In order to reveal the structure of the Hessian, we define the set of $2N-2$
vectors: $v_{p}(\beta^{*}),w_{p}(\beta^{*})$ for $p=1,\dots,i-1,i+1,\dots,N$. 
Each of the vectors is of length $M$ with the $j$-th component being
\begin{subequations}
\begin{align}
[v_{p}(\beta^{*})]_{j} &=  \frac{\sqrt{2}}{\hbar}\int_{0}^{T} \re [\langle
p|U^{\dag}(t,0)H_{\beta^{*}_{j}}(\betastar,t)U(t,0)|i\rangle] \ dt \\
[w_{p}(\beta^{*})]_{j} &=  \frac{\sqrt{2}}{\hbar}\int_{0}^{T} \im [\langle
p|U^{\dag}(t,0)H_{\beta^{*}_{j}}(\betastar,t)U(t,0)|i\rangle] \ dt.
\end{align}
\end{subequations} 
The Hessian matrix with elements in Eq. \ref{hess:elements} may now be written
explicitly in terms of these vectors as 
\begin{equation}\label{hess:top:rank}
\Hessp(\beta^{*}) = -\sum_{p\neq i} [v_{p}(\beta^{*}) \cdot
v_{p}^{\trans}(\beta^{*}) + w_{p}(\beta^{*}) \cdot w_{p}^{\trans}(\beta^{*})].
\end{equation}
Equation \ref{hess:top:rank} shows that the Hessian is a sum of $2N-2$ rank one
matrices.  If the set $\{ v_{p}(\beta^{*}),w_{p}(\beta^{*})\}_{p\neq i}$ is
linearly independent, then the Hessian will have rank $2N-2$; otherwise the rank
is less than $2N-2$. This result has important practical consequences for
optimal control, as shown in the following subsections.

\subsubsection{Robustness at the Top of the Landscape}\label{sec:hess:robust}
A physically attractive control $\beta^{*}$ is one that consistently produces
high yields even in the presence of some degree of noise around $\beta^{*}$ at
or near the top of the landscape.  
In the general context considered here noise may correspond to statistical
variation in the field and/or uncertainty in the time independent Hamiltonian
structure.  
Below, we show that the trace of the Hessian may be utilized as a scalar measure
of the deviation of $\pif$ around a control $\beta^{*}$.  

In order to analyze the behavior at the top of the landscape, consider an
arbitrary variation $d\beta$ about $\beta^{*}$ expressed as
\begin{equation}\label{noise:any}
 d\beta = \sum_{j=1}^{M} db_{j} u_{j}(\beta^{*})
\end{equation}
where $\{ db_{j} \}_{j=1,\dots,M}$ are real expansion coefficients and the set of
$u_{j}(\beta^{*})$ for $j=1,\dots M$ are the eigenvectors of $\Hessp(\beta^{*})$
satisfying 
\begin{equation}
\Hessp(\beta^{*})\cdot u_{j}(\beta^{*}) = \sigma_{j} u_{j}(\beta^{*}), \quad
j=1,\dots,M.
\end{equation}
Importantly, only the first $2N-2$ of the eigenvalues $\sigma_{j}$ are non-zero 
\cite{ash, taksan:control:easy, rabitz:top:pif}.  Here we assume the likely circumstance that
$M \geq 2N-2$; violation of the latter criteria will limit free variations on the landscape
possibly leading to \textit{false} traps on the otherwise nominally trap-free landscape.
Practical circumstances will likely entail $M\approx N^{2}$ easily satisfying $M \geq 2N-2$.
Equation \ref{noise:any} utilizes the fact that the
eigenvectors of the Hessian form a complete set in the $M$-dimensional space of
controls and the expansion coefficients in Eq. \ref{noise:any} may be
interpreted as the projection of the control noise along each eigen-direction of
the Hessian.  
The norm squared of $d\beta$ is given by
\begin{equation}\label{norm:dbeta}
 ||d\beta||^{2} \equiv (d\beta)^{\trans}d\beta = \sum_{j=1}^{M} (db_{j})^{2}.
\end{equation}
The effect of a control variation $d\beta$ in Eq. \ref{noise:any} is given by
Eq. \ref{eq:sec:exp} 
\begin{align}
 \Delta\pif(\beta^{*}) &= (1/2)\sum_{j=1}^{M} \sigma_{j} \ (db_{j})^{2} \\
&= (1/2)\sum_{j=1}^{2N-2} \sigma_{j} \ (db_{j})^{2} \label{sig}
\end{align}
where we have utilized the fact that only the first $2N-2$ eigenvalues of the
Hessian are non-zero \cite{taksan:control:easy, rabitz:top:pif}.  In order to
appreciate the significance of Eq. \ref{sig}, we make the physically reasonable
assumption that the noise is equally distributed along each eigen-direction of
the Hessian such that $(db_{j})^{2} = (db_{k})^{2}$ for all $j,k=1,\dots,M$.  Defining
$||d\beta||^{2} = (d\beta_{0})^{2}$, then it follows from Eq. \ref{norm:dbeta} that
$db_{j} = \pm [(d\beta_{0})^{2}/M]^{1/2}$ and Eq. \ref{sig} becomes 
\begin{align}
 \Delta\pif(\beta^{*}) &= \left( \frac{(d\beta_{0})^{2}}{2M} \right) \textrm{Tr}[
\Hessp(\beta^{*})], \label{delta:M}
\end{align}
since the trace of a matrix is equivalent to the sum of its eigenvalues. 
Utilizing Eq. \ref{hess:top:rank} the deviation in
Eq. \ref{delta:M} can be re-written 
\begin{equation}\label{DeltaPIF}
 \Delta\pif(\beta^{*}) = -\left( \frac{(d\beta_{0})^{2}}{M\hbar^{2}}\right) 
\sum_{p\neq i} \sum_{j=1}^{M} \left| \int_{0}^{T} \langle
p|U^{\dag}(t,0)H_{\beta^{*}_{j}}(\betastar,t)U(t,0)|i\rangle \ dt \right|^{2}.
\end{equation}
Equations \ref{delta:M} or \ref{DeltaPIF} show the relationship between the transition probability
deviation, $\Delta\pif(\beta^{*})$, at the top of the landscape and the Hessian
trace there.  In the worst case scenario, the disturbance $d\beta$ would lie
entirely along the set of $2N-2$ eigenvectors corresponding to non-zero
eigenvalues.  In this circumstance, the factor $M$ appearing in Eq.
\ref{delta:M} would be replaced by $2N-2$, generally increasing the magnitude of
$\Delta\pif(\beta^{*})$.  
In the present work dimensionless units are used for all variables, thereby
avoiding the need for any specific normalization of the Hessian elements.  
Importantly, the response $\Delta\pif$ is invariant to such a
normalization. 
\subsubsection{Hessian at the Bottom of the Landscape}\label{sec:hess:bott}
Although the bottom of the landscape is not a desirable location, behavior there
is important to explore as initial controls will likely result in $\pif$ having
a small value.  Here, we take this to the limit and explore level sets at the
absolute bottom corresponding to controls $\beta^{*}$ producing $\pif=0.0$.  At
the bottom of the landscape, $\langle i|U^{\dag}(T,0)|f\rangle = 0$, utilizing
Eqs. 4 and 5 significantly simplifies Eq. \ref{second:gen} to the form
\begin{equation}
\begin{split}
\frac{\partial^{2}\pif}{\partial\beta^{*}_{j}\partial\beta^{*}_{k}}
& =(-2/\hbar^{2}) \re \Big[ \int_{0}^{T} \langle q| 
U^{\dag}(t)H_{\beta^{*}_{j}}(\betastar,t)U(t)|i\rangle dt  \\
&\quad\quad \times \int_{0}^{T} \langle i| U^{\dag}(t) H_{\beta^{*}_{k}}(\betastar,t) U(t)|q\rangle \
dt \Big] 
\end{split}
\end{equation}
where $|q\rangle \equiv U^{\dag}(T,0)|f\rangle$.  
As for the top of the landscape in Section 2.2.1, once again the Hessian rank at
the bottom is important to assess.  In order to do so,  we define the two
vectors $y(\beta^{*})$ and $z(\beta^{*})$ of length $M$ with entries 
\begin{subequations}
\begin{align}
[y(\beta^{*})]_{j} &=  \frac{\sqrt{2}}{\hbar}\int_{0}^{T} \re [\langle
q|U^{\dag}(t,0)H_{\beta^{*}_{j}}(\betastar,t)U(t,0)|i\rangle] \ dt \\
[z(\beta^{*})]_{j} &=  \frac{\sqrt{2}}{\hbar}\int_{0}^{T} \im [\langle
q|U^{\dag}(t,0)H_{\beta^{*}_{j}}(\betastar,t)U(t,0)|i\rangle] \ dt.
\end{align}
\end{subequations} 
Then, the Hessian at the bottom may be written explicitly as 
\begin{equation}\label{hess:bott}
\Hessp(\beta^{*}) = [y(\beta^{*}) \cdot y^{\trans}(\beta^{*}) + z(\beta^{*})
\cdot z^{\trans}(\beta^{*})]
\end{equation}
and utilizing Eq. \ref{hess:bott}, the Hessian trace may be expressed as
\begin{equation}\label{hess:bott:trace}
 \textrm{Tr}[\Hessp(\beta^{*})] = 
(2/\hbar^{2}) \sum_{j=1}^{M} \left| \int_{0}^{T} \langle q| U^{\dag}(t,0)H_{\beta^{*}_{j}}(\betastar,t)U(t,0)|i\rangle \ dt \right|^{2}.
\end{equation}
Similar to the discussion in Section \ref{sec:hess:top} regarding Eq.
\ref{hess:top:rank}, now Eq. \ref{hess:bott} shows that the rank of the Hessian
at the bottom is at most 2.  The maximal case of rank $2$ occurs if the vectors
$y(\beta^{*})$ and $z(\beta^{*})$ are linearly independent;  otherwise, the rank
of the Hessian at the bottom can be less than 2.  
Thus, at most there are two coordinated variations around the control
$\beta^{*}$ that will lift $\pif(\beta^{*})$ off the bottom, as discussed below.

\subsubsection{Climbing Rapidly from the Bottom of the
Landscape}\label{sec:bott:climb}
At the top of the landscape Section 2.2.2 showed that the Hessian trace is an
indicator of the robustness of a control solution.  We now similarly show how
the Hessian trace may influence gradient climbs originating from controls
producing $\pif$ values near the landscape bottom.
To exhibit the influence of the Hessian trace on gradient based searches
consider a control $\beta'$ near a control $\beta^{*}$ at the bottom of the
landscape.  Here we assume that $\beta'$ corresponds to a point slightly off the bottom of the landscape such that $\pif(\beta') > 0.0$. 
The gradient at $\beta' = \beta^{*} + d \beta$ can be written as
\begin{align}
 \nabla\pif(\beta') &= \nabla\pif(\beta^{*}+d \beta) \nonumber \\
& \approx \Hessp(\beta^{*}) \cdot d\beta. \label{hess:act:alpha}
\end{align}
In the present context, $d\beta$ could arise either due to noise or considered
as a concerted control variation.
The gradient norm squared is
\begin{align}
 K(\beta') &\equiv || \nabla\pif(\beta')||^{2} \nonumber \\ 
&\approx || \Hessp(\beta^{*}) \cdot d \beta ||^{2} \label{eq:here}.
\end{align}
At the point $\beta^{*}$ at the landscape bottom, the Hessian can be rewritten
as
\begin{equation}
 \Hessp(\beta^{*}) = \sigma_{1} \ u_{1}(\beta^{*}) \ u_{1}(\beta^{*})^{\trans} +
\sigma_{2} \ u_{2}(\beta^{*}) \ u_{2}(\beta^{*})^{\trans} 
\end{equation}
where $\sigma_{1}$ and $\sigma_{2}$ are the two non-zero positive eigenvalues of
the Hessian with $u_{1}(\beta)$ and $u_{2}(\beta)$ being the corresponding
eigenvectors.  Then, Eq. \ref{eq:here} becomes
\begin{equation}\label{eq:nonzero}
 K(\beta') = \sigma^{2}_{1} (d \beta^{\trans} \cdot u_{1}(\beta^{*}))^{2} + 
 \sigma^{2}_{2} (d \beta^{\trans}\cdot u_{2}(\beta^{*}))^{2}.
\end{equation}
In order to physically interpret Eq. \ref{eq:nonzero} it is convenient to
consider a disturbance of unit magnitude $||d \beta|| = 1$.  If $d \beta$ lies
entirely in the control space spanned by the two Hessian eigen-directions
$u_{1}$ and $u_{2}$, then the gradient norm is bounded by $\sigma_{1}$ and
$\sigma_{2}$.  However, in the more likely scenario that the vector $d \beta$
also has a component lying in the null space of the Hessian, then the gradient
norm is bounded 
\begin{equation}\label{grad:bounds}
 0 \leq ||\nabla\pif(\beta')||\leq \max\{\sigma_{1},\sigma_{2}\},\quad ||d
\beta||=1.
\end{equation}
Although maximizing the trace of the Hessian does not necessarily imply a larger
upper bound for the gradient norm in Eq. \ref{grad:bounds}, experience from numerical
simulations (not shown) reveals that strongly curved portions of the landscape bottom generally enable a rapid
rise in $\pif$ when aiming at ascent.  

\section{Algorithms for Landscape Explorations}
Previous formal work has shown that level sets of controls exist for all values
of $\pif$ including at the top and bottom of the landscape
\cite{rabitz:quant:opt, rabitz:top:pif, taksan:control:easy}.  Each level set
member produces the same transition probability value; however, controls over
the level set can show a large variation of secondary characteristics.   These
characteristics may include the mechanism by which control is achieved, the
stability of each solution to disturbances, etc.  Here, a D-MORPH algorithm is
introduced to identify level set members aimed at locating control solutions
displaying secondary desirable characteristics in addition to achieving
prescribed values of $\pif$.  The myopic nature of the D-MORPH algorithm
enables the exploration of fundamental landscape issues, as well as provides a
link to laboratory landscape excursions.  The D-MORPH method has already been
implemented in the laboratory where it has been shown to work at the top and
bottom of the landscape \cite{roslund2008, roslund:thesis}.
We are interested in finding solutions $\beta^{*}$ that belong to a level set
denoted as $\level$, corresponding to either $\pif(\beta^{*})=1.0$ or
$\pif(\beta^{*})=0.0$. In order to explore the level set, $\level$, the control
$\beta^{*}$ is parametrized with a continuous parameter $s$, i.e.,
$\beta^{*}\Rightarrow \beta^{*}(s)$.  This parameterization can be viewed as a
curve $\beta^{*}(s)$ as a function of $s$ through control space. The initial
control $\beta(s=0)$ is not likely to lie on a level set $\level$ calling
for an ascent or descent of the landscape.

The goal of the next section is to derive differential equations for $\beta(s)$ of the form
\begin{equation}\label{gen:dmorph}
 d\beta(s)/ds = F[\beta(s),s]. 
\end{equation}
A particular form of Eq. \ref{gen:dmorph} will produce a trajectory that can
ascend/descend to the landscape top/bottom starting from an initial control
$\beta$ while another form of Eq. \ref{gen:dmorph} will produce a
\textit{trajectory} $\beta^{*}(s) \in\level$ that lies on the level set
maintaining either $\pif(\beta^{*}(s))=1.0$ or $\pif(\beta^{*}(s))=0.0$ upon starting
from an initial point at the respective landscape extremum.  Section \ref{dmorph:climb} presents the
differential equation for ascending or descending the landscape and Section
\ref{dmorph:secondorder} presents the algorithm for exploring the level sets
at values of $\pif=1.0$ and $\pif=0.0$. Section \ref{free} shows that the
D-MORPH algorithm may be utilized to locate solutions exhibiting secondary
desirable physical characteristics along level sets at either $\pif=1.0$ or $0.0$. 

\subsection{Ascending and Descending the Landscape}\label{dmorph:climb}
Exploring level sets at the top or bottom of the landscape first necessitates
finding an initial control residing at either location.  
For this purpose, the D-MORPH technique can be used to ascend or descend the
landscape from an arbitrary trial control $\beta(0)$.  Differentiating $\pif(s) =
\pif(\beta(s))$ with respect to $s$ produces
\begin{equation}\label{1st:dmorph:eq}
d\pif(\beta(s))/ds = \gradp(\beta(s))^{\trans}\cdot d\beta(s)/ds.
\end{equation}
Choosing 
\begin{equation}\label{1st:climb}
 \partial \beta(s)/\partial s = \rho(s) \gradp(\beta(s))
\end{equation}
for $\rho(s) > 0$ guarantees ascension since 
$d\pif(s) /ds \geq 0$; while choosing $\rho(s) < 0$ will result in descension. 
The function $\rho(s)$ may be chosen as desired, and its magnitude 
will dictate the rate of ascent or descent.  Equation
\ref{1st:dmorph:eq} will be used to identify controls $\beta^{*}$ residing at
the landscape top and bottom.  

\subsection{Movement on the Top and Bottom Level Sets}\label{dmorph:secondorder}
This section is concerned with movement along the level set of control solutions
either at the bottom or top of the control landscape, corresponding to
$\beta^{*}(s)$ values maintaining either $\pif(\beta^{*}(s))=0.0$ or
$\pif(\beta^{*}(s))=1.0$.
In order to derive an equation for the desired trajectories, we 
take into account that the first derivative $\gradp(\beta^{*})$ is necessarily zero 
at the top or bottom of the landscape and therefore differentiate
$\pif(\beta^{*}(s))$ twice with respect to s, 
\begin{align}\label{dmorph:2nd:ds}
 d^{2}\pif(\beta^{*}(s))/ds^{2} = (1/2)  \left(\partial\beta^{*}(s)/\partial
s\right)^{\trans}\cdot \Hessp(\beta^{*}(s))\cdot
\left(\partial\beta^{*}(s)/\partial s\right).
\end{align}
Investigating the trajectories $\beta^{*}(s)$ over the top or bottom of
the landscape is assured by demanding that $d^{2}\pif(\beta^{*}(s))/ds^{2}=0$. 

The analysis in Sections \ref{sec:hess:top} and \ref{sec:hess:bott}, showed that
the rank of the Hessian is at most $L=2N-2$ and $L=2$ at the top and bottom,
respectively. Since the Hessian $\Hessp$ is a real symmetric $M\times M$ matrix,
it has $M$ real eigenvalues.  The Hessian rank being $L$, with generally $L<M$, means that there are additionally
$M-L$ zero eigenvalues.  Thus, the Hessian can be written as
\begin{equation}\label{hess:eigen}
\Hessp(\beta^{*}(s)) = \sum_{\ell=1}^{L} \sigma_{\ell} u_{\ell}(\beta^{*}(s))
\cdot u_{\ell}^{\trans}(\beta^{*}(s)).
\end{equation}
where the $\{ u_{\ell}(\beta^{*}(s))\}$ are the orthonormal eigenvectors of
$\Hessp(\beta^{*}(s))$ corresponding to the $L$ non-zero eigenvalues, $\sigma_{\ell}$. 

Using Eqs. \ref{dmorph:2nd:ds} and \ref{hess:eigen}, the criterion
$d^{2}\pif(s)/ds^{2}=0$ specifies the following $L$ linear constraints on
$d\beta^{*}(s)/ds$
\begin{equation}\label{constraints}
 u_{\ell}(\beta^{*}(s))^{\trans}\cdot d\beta^{*}(s)/ds = 0 \quad \ell =
1,\dots,L.
\end{equation}
The constraints given by Eq. \ref{constraints} can be re-expressed as a
differential equation whose solution is a trajectory $\beta^{*}(s)$ on the level
set $\level$ at either the top or bottom of the landscape,
\begin{equation}\label{dmorph:diff:2}
d\beta^{*}(s)/ds = \Pcrit(\beta^{*}(s)) \cdot g(\beta^{*}(s),s),
\end{equation}
where $g(\beta^{*}(s),s)$ is a vector of length $M$ that can be freely chosen. 
The projector $\Pcrit$ is specified by its action
\begin{equation}\label{proj:2nd}
\Pcrit(\beta^{*}(s)) \cdot g(\beta^{*}(s),s) \equiv g(\beta^{*}(s),s) - \sum_{\ell}
\ u_{\ell}(\beta^{*}(s)) \ [u_{\ell}(\beta^{*}(s))^{\trans}\cdot
g(\beta^{*}(s),s) ]  
\end{equation}
which removes any component of the vector $g(\beta^{*}(s),s)$ lying in the space
spanned by the set of eigenvectors $\{ u_{\ell}(\beta^{*}(s))\}$.  Thus,
$\Pcrit(\beta^{*}(s))$ is the projector onto the null space of
$\Hessp(\beta^{*}(s))$ thereby assuring that the solution of Eq. \ref{dmorph:diff:2}, $\beta^{*}(s)$,
will satisfy the constraints in Eq. \ref{constraints}; direct substitution of Eq. \ref{dmorph:diff:2} into Eq. \ref{constraints}
upon use of Eq. \ref{proj:2nd} shows that the demand $u_{\ell}(\beta^{*}(s))^{\trans}\cdot d\beta^{*}(s)/ds = 0$ is
satisfied.
The choice of free vector $g(\beta^{*}(s),s)$ will specify the particular trajectory on the
level set.

\subsection{Particular Control Trajectories at the Top or Bottom of the Landscape to Meet Auxiliary
Objectives}\label{free}
The freedom in choosing the vector $g$ in Eq. \ref{dmorph:diff:2} will lead to
different D-MORPH trajectories over the top or bottom level sets.  This freedom allows
for secondary control behavior demands to be placed on level set explorations.  For example,
solutions may be sought that achieve high yield while either minimizing
the effects of parameter disturbances on the overall outcome, minimizing the
pulse fluence or optimizing other physical properties.  

Consider, for example, the goal of moving on the level set $\level$ while also
seeking to maximize a secondary cost function $J(\beta^{*})$, such as robustness
to variations $\beta^{*}\rightarrow \beta^{*}+d\beta$ at the top of the
landscape.  As the trajectory specified by the solution to Eq.
\ref{dmorph:diff:2} advances, changes in $J$ will be given by
\begin{equation}\label{dJds}
dJ(s)/ds = \gradJ(\beta^{*}(s))^{\trans}\cdot d\beta^{*}(s)/ds,
\end{equation}
and substituting Eq. \ref{dmorph:diff:2} into Eq. \ref{dJds} gives
\begin{equation}
dJ(s)/ds = \gradJ(\beta^{*}(s))^{\trans}\cdot \Pcrit(\beta^{*}(s))\cdot
g(\beta^{*}(s),s).
\end{equation}
As the projector $\Pcrit(\beta^{*}(s))$ is positive semi-definite, choosing
$g(\beta^{*}(s),s) = \rho'(s)  \gradJ(\beta^{*}(s))$,
with $\rho'(s) > 0$ will locally maximize $J$ over the trajectory constrained to the
level set $\level$ and the value of $\rho'(s)$ will mediate the rate that $J$ rises
in the process.  Similarly, choosing $\rho'(s) < 0$ will lead to minimization of $J$ over
the level set $\level$.  

\section{Illustrations}
Exploring the level sets at the top or bottom necessitates first either
ascending or descending the landscape with Eq. \ref{1st:climb}.  Upon
reaching either $\pif=1.0$ or $\pif=0.0$ (within adequate tolerance), we then move
over the associated level set by solving the second-order D-MORPH equation in
Eq. \ref{dmorph:diff:2}. The D-MORPH ascent/descent Eq. \ref{1st:climb} is
solved using a Runge-Kutta fourth-order variable step size integrator over the
$s$-domain.  
The second-order D-MORPH Eq. \ref{dmorph:diff:2} is solved using the same
integrator until $\pif$ deviates beyond a specified tolerance, $\delta$, from
the level set value. In the following examples, the \textit{top} of the
landscape is defined as $\pif > 1-\delta $ and the \textit{bottom} of the
landscape as $\pif < \delta$ where $\delta = 10^{-8}$.  If the tolerance is
violated, a switch to the first order ascent/descent Eq. \ref{1st:climb} is
made to, respectively, ascend or descend back to the level set at the top or
bottom. 

The examples explore the top and bottom of the landscape for Hamiltonians  of
the form
\begin{equation}\label{ham_form}
 H(t) = H_{0}-\mu\epsilon(t).
\end{equation}
where $H_{0}$ is diagonal, $\mu$ is a real symmetric dipole matrix and
$\epsilon(t)$ is the electric control field, 
\begin{equation}\label{field_expand}
 \epsilon(t) = S(t) \sum_{ij}^{P} a_{ij} \sin(\omega_{ij}t + \phi_{ij})
\end{equation}
with $P$ contributing frequencies and the envelope
\begin{equation}
S(t) = \exp\left\{-\frac{(t-T/2)^{2}}{2\sigma^{2}}\right\}.
\end{equation}
The indices '$ij$' in Eq. \ref{field_expand} correspond to the system transition
frequencies and are explicitly stated for each particular illustration below.  We set
$\sigma = 3.5$ in the simulations.  The control variables $\beta$ are drawn from
(a) the set of Hamiltonian structure parameters $\{ E_{j},\mu_{j},\mu_{ij}\}$ where
$E_{j}$ is the $j$-th system energy level and $\mu_{j}$ and $\mu_{ij}$ are 
diagonal and off-diagonal dipole matrix elements, respectively, and (b) the
field parameters $\{a_{ij}, \phi_{ij} \}$ where $a_{ij}$ is the amplitude and
$\phi_{ij}$ the phase associated with the $\omega_{ij}$ spectral component.  The
goal is to explore controls at the top or bottom of the landscape through the
Hamiltonian structure variables of $H_{0}$ and $\mu$ in Eq. \ref{ham_form}, and/or the applied field
through the amplitudes and phases in Eq. \ref{field_expand}.   
In these simulations, dimensionless units are used, $\hbar$ is set to
1.0, and time is discretized between $t=0$ and $T=20$ into $4096$ points.  
The selected cases below are chosen to illustrate basic physical behavior at the top and bottom of the landscape 
that was identified in an extensive set of simulations.

\subsection{Level Sets at the Top of the Hamiltonian Structure Control
Landscape}
Here, we consider a simple $N=2$ state system with $M=3$ Hamiltonian structure
control variables 
$E_{2}, \mu_{2}$ and $\mu_{12}$ along with a fixed electric field $\epsilon(t)$.
 This simple system is chosen for investigation as it permits a clear picture of
level set exploration at the top of the landscape.  

For illustration, seven distinct initial systems specified by the values of
$E_{2},\mu_{2}$ and $\mu_{12}$ were generated with each being a point on the top
of the landscape.  The fixed field, $\epsilon(t)$ was chosen using the form in
Eq. \ref{field_expand} with $P=4$, and $\epsilon(t)$ merely serves the role of
facilitating utilization of the controls $E_{2},\mu_{2}$ and $\mu_{12}$.  The
field parameters are given in \cite{field_strands}.  As initial systems, we
chose $E_{2} = \mu_{2} = 0$ and a simple search over $\mu_{12}$ was performed to find
that $\mu_{12} = \{ 0.14,0.42,0.70,0.98,1.26,1.54,1.82\}$ satisfy $\pif=1.0$
\cite{ash}.  Following the discussion in Section \ref{sec:hess:top}, the rank of
the Hessian at the top of the landscape was confirmed numerically to be $2N-2 =
2$.   Thus, with three controls, there is an underlying one dimensional
level set satisfying $P_{1\rightarrow 2} = 1.0$; the level set is a curve
$\{E_{2}(s), \mu_{2}(s), \mu_{12}(s)\}$ parameterized by $s$ lying in the three dimensional control space. 

In the simulations, the free vector $g(\beta^{*}(s),s)$ in Eq.
\ref{dmorph:diff:2} was chosen to be the eigenvector $u^{0}(\beta^{*}(s))$
corresponding to the only non-zero eigenvalue of the Hessian.  Since the
eigenvectors of the Hessian are orthogonal, this choice of free vector
corresponds to simply moving in the direction given by $u^{0}(\beta^{*}(s))$.  
The integration of Eq. \ref{dmorph:diff:2} was performed over the domain $s\in[0,20]$. 
Starting from the initial points $\{E_{2}(0), \mu_{2}(0), \mu_{12}(0)\}$ above, 
Fig. 2 shows the seven one dimensional trajectories with each tracing out a path
on the level set at the top of the landscape, $P_{1\to 2} = 1.0$.  Each level set
trajectory is seen to be a highly coordinated path through the space of controls
$E_{2}, \mu_{2}$ and $\mu_{12}$.  Although it is possible that the seven paths
actually are segments of a single overall level set trajectory (i.e., the paths
eventually join together) this circumstance was not found to be the case over
the domain explored here.  Thus, 
Fig. 2 also illustrates that level sets \textit{may} have disconnected
components.  The highly contorted nature of the level set trajectories in Fig. 2 illustrates
the complex structure of the top of the landscape.

\subsection{Robustness at the Top of the Combined Hamiltonian Structure and
Field Control Landscape}
In this section, we explore the robustness of an $N=5$ level system with fixed
diagonal $H_{0}$ energies $E_{1} = 0, E_{2} = 1, E_{3} = 4, E_{4} = 9,
E_{5}=16$.  There are $M=12$ total controls $\beta^{*}$ drawn from: the
dipole matrix elements, $\{ \mu_{j,j+1}\}_{j=1,\dots,4}$ as well as
the field amplitudes $\{ a_{j,j+1} \}_{j=1,\dots,4}$ and phases $\{
\phi_{j,j+1}\}_{j=1,\dots,4}$.  The remaining dipole elements and field amplitudes and phases
were fixed at zero.  The frequency associated with the $j \rightarrow j+1$ transition
is given by $\omega_{j,j+1} = |E_{j+1}-E_{j}|$ and the field had the form in Eq. \ref{field_expand}
with $P=4$.

In this illustration, the trajectory of controls $\beta^{*}(s)$ was determined
with an extra cost as the trace of the Hessian, 
\begin{equation}
   J = \textrm{Tr}(\Hessp) = \sum_{j=1}^{12} \frac{\partial^{2}\pif}{\partial\beta^{*2}_{j}}
=  \sum_{\ell=1}^{8} \sigma_{\ell} < 0,
 \end{equation}
introduced to explore the
curvature of the level set at the top of the landscape, $\pp=1.0$. 
The value of $J$ directly influences the robustness to variations in the controls $d\beta$ as
evident from $J$ being the sum of the Hessian eigenvalues, $\sigma_{\ell}$.
An initial
control at the top of the landscape was found through Eq. \ref{1st:climb}
with $\rho(s)=1.0$ starting from the initial set of four amplitudes, phases and
dipoles given in \cite{param_top_ex}.  The ascent procedure converged to the
nominal control value $\beta_{0}^{*}$ denoting the amplitudes, phases and dipole
values producing $P_{1\rightarrow 5} \simeq 1.0 $.  In the ascent
$J$ was not explicitly considered, but its value of $J=-180.9$ was recorded at
$\beta_{0}^{*}$ as a reference for comparison.  The nominal control variables,
$\beta_{0}^{*}$, were used as the initial condition for a calculation designed
to locate the most robust solution, $\beta_{\textrm{rob}}^{*}$ at the top of the landscape.  
Following the logic in Section \ref{free}, the free vector in Eq. \ref{dmorph:diff:2} was
chosen as $g_{\textrm{rob}}(\beta^{*}(s))=\nabla J(\beta^{*}(s))$, i.e.,
$\beta^{*}_{\textrm{rob}}$ corresponds to the smallest attained value of $|J|$,
as $J<0$.  This choice of free vector follows from the discussion of
robustness in Section \ref{sec:hess:robust}. The gradient $\nabla
J(\beta^{*}(s))$ was calculated using a finite difference scheme.  The solution
$\beta_{0}^{*}$, producing $J=-180.9$ along with the solution
$\beta_{\textrm{rob}}^{*}$ producing $J=-30.3$ are shown in Fig. 3a and 3b.,
respectively. 

To investigate the robustness of the control $\beta^{*}$ at the landscape top,
the Hessian matrix for the solutions $\beta_{0}^{*}$ and
$\beta_{\textrm{rob}}^{*}$ was calculated with Eq. \ref{hess:top:rank}.  The
Hessian eigenvalue of largest magnitude was $-69.7$ for $\beta_{0}^{*}$ while
the corresponding value was $-14.1$ for $\beta_{\textrm{rob}}^{*}$.  The
eigenvector associated with the largest magnitude eigenvalue for both solutions
is plotted in Fig 4.  Fig. 4a shows that the control solution $\beta_{0}^{*}$ is most
sensitive to decreasing the magnitude of the dipoles $\mu_{12}$ and $\mu_{23}$.
In contrast, Fig. 4b shows that the control solution $\beta_{\textrm{rob}}^{*}$
is most sensitive to decreasing the magnitude of the amplitude $a_{45}$ and the
dipole $\mu_{45}$ (the nominal value of $\mu_{45} < 0$ in Fig. 3b) as well as increasing the magnitude of
$a_{34}$ and $\mu_{34}$.    

Together, Figs. 3 and 4 show the manner in which a more robust solution was
established.  Over the level set trajectory (i.e., on going from Fig. 3a to Fig.
3b), the magnitude of the field amplitudes were decreased by $\sim 70\%$
while the magnitude of the dipoles were increased by approximately the same
amount.  This \textit{redistribution} of parameter magnitude produces a more
stable solution at the top of the landscape.  The small magnitude of the dipole
matrix elements in Fig 3a lead to very unstable solutions when amplitude, phase
and dipole matrix element noise is introduced into the system.  This is evident
from the large eigenvector contribution of $\mu_{12}$ and $\mu_{23}$ in Fig. 4a.
 Conversely, the balanced distribution of about equal magnitude for the field
amplitudes and dipole elements in Fig 3b provides a more stable solution. 
Although the changes in the phases are more subtle to interpret, the
optimization process manipulated these as well in order to gain robustness.

In order to assess the degree of robustness of the control solutions, a
simulation was run to test the effect of random disturbances around
$\beta_{0}^{*}$ and $\beta_{\textrm{rob}}^{*}$.   
Disturbances were introduced by selecting a random vector, $p_{\textrm{rand}}$,
where each entry was chosen from a normal distribution with zero mean and
standard deviation of $1.0$.  Then $p_{\textrm{rand}}$ was normalized to produce
$||p_{\textrm{rand}}|| = 0.25$.  For $5000$ runs, the control solutions
$\beta_{0}^{*}$ and $\beta_{\textrm{rob}}^{*}$ were perturbed by the same random
vector, e.g., $\beta' = \beta^{*} + p_{\textrm{rand}}$, and $P_{1\rightarrow
5}(\beta')$ was recorded.  The results are given in Fig. 5 where panel (a)
corresponds to $\beta_{0}^{*}$ and panel (b) corresponds to
$\beta_{\textrm{rob}}^{*}$.  

Figure 5 shows that the nominal control solution in panel (a) is highly
sensitive to the perturbations introduced by $p_{\textrm{rand}}$ while the
robust control solution in panel (b) is minimally sensitive to the random
perturbations introduced by $p_{\textrm{rand}}$.  Although both control
solutions are able to achieve a very high yield, $P_{1\rightarrow 5} \simeq 1.0$, only the solution in panel (b)
can tolerate significant disturbances in the parameter settings.  The mean value
of $P_{1\rightarrow 5}$ for panel (a) is $0.612$ with a left standard deviation
of $0.174$ while the mean value of $P_{1\rightarrow 5}$ for panel (b) is $0.936$
with a left standard deviation of $0.040$.  Additional tests of robustness were
performed on these control solutions including stability to relative
disturbances, i.e., $\beta' = \beta^{*}(1+p_{\textrm{rand}})$ that confirmed the
robustness of $\beta_{\textrm{rob}}^{*}$ over that of $\beta_{0}^{*}$ (not
shown) but to a lesser degree.

\subsection{The Ease of Climbing off the Bottom of the Control Field Landscape}
The bottom of the landscape is a domain where the goal is to ascend as rapidly
as possible.  As $\gradp=0$ at the bottom, the curvature dictates the rate of
climbing.  Thus, in this example, we investigate the curvature extremes
encountered at the landscape bottom for the target $1\rightarrow 3$ transition
by analyzing
the trace of the Hessian over families of control solutions producing $P_{1
\rightarrow 3} = 0.0$.  
An $N=4$ state system is considered with fixed $H_{0}$ and $\mu$ specified by
$E_{1} = 0.0, E_{2} = 1.0, E_{3}=4.0, E_{4} = 9.0, \mu_{12} = 0.50, \mu_{13} =
0.25, \mu_{14} = 0.15, \mu_{23} = 0.20, \mu_{24} = 0.10, \mu_{34} = 0.05$, and
the control field for variation has the form in Eq. \ref{field_expand} with $P=6$ corresponding
to $M=12$ amplitude and phase control variables.  An
initial control solution, $\beta_{0}^{*}$, was found producing $P_{1\rightarrow
3} = 0.0$ using the descent procedure in Eq. \ref{1st:climb} with $\rho(s) =
-1.0$, and two separate D-MORPH runs were initiated at $\beta_{0}^{*}$ with the
goals of
either maximizing or minimizing the Hessian trace, $J =
\textrm{Tr}[\nabla^{2}\pif]$, in Eq. \ref{hess:bott:trace} while remaining at the
bottom of the landscape.  Here, the free vector in Eq. \ref{dmorph:diff:2} is
chosen as $g=\pm \gradJ$, with $+$ for maximization and $-$ for minimization of
the Hessian trace. 
The control field solution that maximized the Hessian trace produced
$\beta_{\textrm{max}}^{*}$ with $J=8\times 10^{3}$, while the solution that
minimized the Hessian trace produced $\beta_{\textrm{min}}^{*}$ with $J=0.31$.

Figure 6 shows the dynamics of state $|3\rangle$ induced by both control
solutions where the value $P_{1\rightarrow 3} \simeq 0.0$ is reached at the final
time.  Contrary to what might be expected, the population of state $|3\rangle$
over time is non-trivial at the landscape bottom with dynamics varying greatly
over the breadth of the level set.  
In particular, $P_{1\rightarrow 3}(t)$ for the maximal trace control solution, $\beta_{\textrm{max}}^{*}$, 
significantly lifts off the bottom at intermediate times to only finally return
to $P_{1\rightarrow 3}\simeq 0.0$ at the final target time.  In contrast, the
minimal trace control solution, $\beta_{\textrm{min}}^{*}$, has $P_{1\rightarrow 3}(t) \simeq 0.0$ over the
full time interval $0 \leq t \leq 20$.  Moreover, from these simulations, the
mean Hessian eigenvalue at the most curved portion of the level set, corresponding to  $\beta_{\textrm{max}}^{*}$,
is found to be more than three orders of magnitude larger than the mean eigenvalue at the flattest
portion of the level set corresponding to $\beta_{\textrm{min}}^{*}$.

To investigate the influence of the curvature at the landscape bottom on
attempts to climb from there ($\pif \approx 10^{-8}$), two
gradient ascents were initiated, respectively, originating from the maximal and
minimal trace producing fields.  These were done over the interval $s\in[0,2]$
using Eq. \ref{1st:climb} with $\rho(s) = 1$ and the results are shown in Fig. 7.  
The gradient ascent starting at the maximal trace field $\beta^{*}_{\textrm{max}}$ (corresponding to the left-ordinate)
reaches a maximum value of $P_{1\to 3} \approx 0.999$ at the final time and $s=2$ while
the gradient ascent starting at the minimal trace field
$\beta^{*}_{\textrm{min}}$ (corresponding to the right-ordinate) reaches a
maximum value of $P_{1\to 3} \approx 9.5 \times 10^{-9}$ at the final time and $s=2$. 
Here, and in a large number of other simulations, it is evident that the trace
of the Hessian at the bottom of the control landscape has a significant
influence on the rapidity with which searches can ascend from there. On
the abscissa scale in Fig. 7, the optimization starting from the minimal trace
field $\beta^{*}_{\textrm{min}}$ reaches a value of $P_{1\to 3} > 0.9$ only at
$s\approx 68$.           

These results on the varying degrees of curvature at the landscape bottom are
significant because
most searches for effective control fields start out with low yields.  This
circumstance is especially important for complex systems where no intuitive set
of good trial controls may exist to begin the optimization procedure.  
However, these simulations also show that there can be attractive classes of
trial controls corresponding to extremely curved portions of the control
landscape at the bottom.  

\section{Conclusion}

This work considered the transition probability landscape as a function of
either the Hamiltonian structure and/or an applied field.  Level sets were shown
to exist at the landscape top $\pif=1.0$ and bottom $\pif=0.0$.  Each level set can
contain control solutions displaying a range of secondary characteristics,
including robustness to disturbances at the top and rapid climbing capability from the
bottom.  We also introduced an extended version of the D-MORPH algorithm for 
identifying desirable control solutions at the landscape top or
bottom.  This advance complements the prior algorithm \cite{rothman:dmorph}
for exploring transition probability values over the interior domain $0 < \pif <
1.0$.  Taken together, these D-MORPH algorithms provide the means to explore the
control landscape and its level sets at any value of $\pif$ to gain fundamental
insights as well as for the practical concerns of finding robust control
solutions.  

Although the D-MORPH procedure was developed here in a theoretical 
and computational context, it is also possible to extend some of the core
concepts to the laboratory 
where statistical sampling of the controls \cite{roslund2008} can be used to
obtain first (gradient) and second-order (Hessian) information (i.e., either by
varying the applied field or Hamiltonian structure).  
In particular, this procedure has been used to measure the Hessian at the top
and bottom of the $N=4$ level atomic Rubidium landscape and travel along the
associated level sets \cite{roslund:thesis}.
A general procedure might implement a closed-loop technique coupled to a
learning algorithm capable of \textit{updating} the field and Hamiltonian
parameters progressively towards an optimal set at the top or bottom
\cite{judson:rabitz} especially for more complex systems.  

The ability to explore ancillary goals (e.g., robustness to field noise or
Hamiltonian structure variations illustrated here) at the highest (and lowest) 
yields reveals additional glimpses of the fundamental control landscape.  Other
ancillary goals can be envisioned
including those that place demands on the nature of the mechanism that achieves
the control goal.  For example, 
if the target is controlled rearrangement of a molecule then a cost can be
placed against populating any states that
lead to dissociation or other undesirable physical processes.  Although this
study focused on the transition
probability landscape, the concepts here can be applied to the quantum control
landscape for any observable.
Furthermore, these tools can be extended to address an ensemble of systems using
the density matrix formalism.  

\section*{Acknowledgements}
The authors acknowledge funding from the DOE and ARO.  J.D. acknowledges funding
from the Program in Plasma Science and Technology.

\clearpage
\bibliographystyle{ieeetr}

\begin{thebibliography}{10}
\bibitem{field_strands}
Since the system energies are permitted to vary, there is no natural labeling
  of frequencies. Here, we just list the frequencies chosen, along with the
  corresponding amplitudes and phases. Frequencies = $\{ 1.10, 0.16, 0.07,
  1.00\}$. Amplitudes = $\{ 1.61, 2.25, 0.74, 2.07\}$. Phases = $\{ -1.15,
  -0.11, -1.71, -0.46\}$.
\bibitem{param_top_ex}
The initial set of amplitudes, phases and dipoles produced $\pif \approx 0.0$
  and were given by: $a_{12} = 1.80, a_{23} = 2.94, a_{34} = 1.33, a_{45} =
  1.38, \phi_{12} = 1.65, \phi_{23} = 0.88, \phi_{34} = 2.95, \phi_{45} = 3.09,
  \mu_{12} = -1.00, \mu_{23} = -0.72, \mu_{34} = -0.35, \mu_{34} = 0.05$.
\end{thebibliography}

\clearpage

\begin{figure}
 \centering
 \includegraphics[scale=1.5]{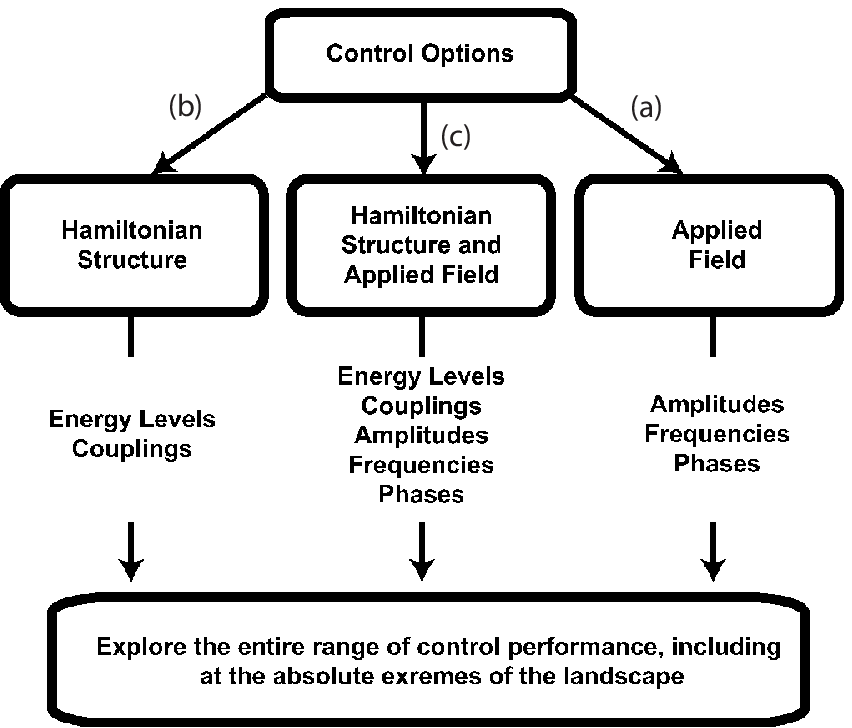}
\newline\newline Figure 1
\end{figure}
\clearpage

\begin{figure}
 \centering
 \includegraphics[scale=1.5]{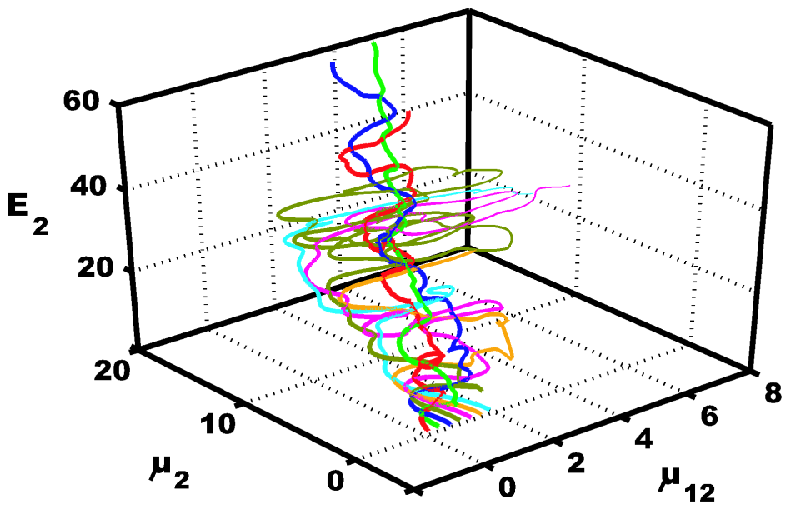}
\newline\newline Figure 2
\end{figure}
\clearpage

\begin{figure}
 \centering
 \includegraphics[scale=1.00]{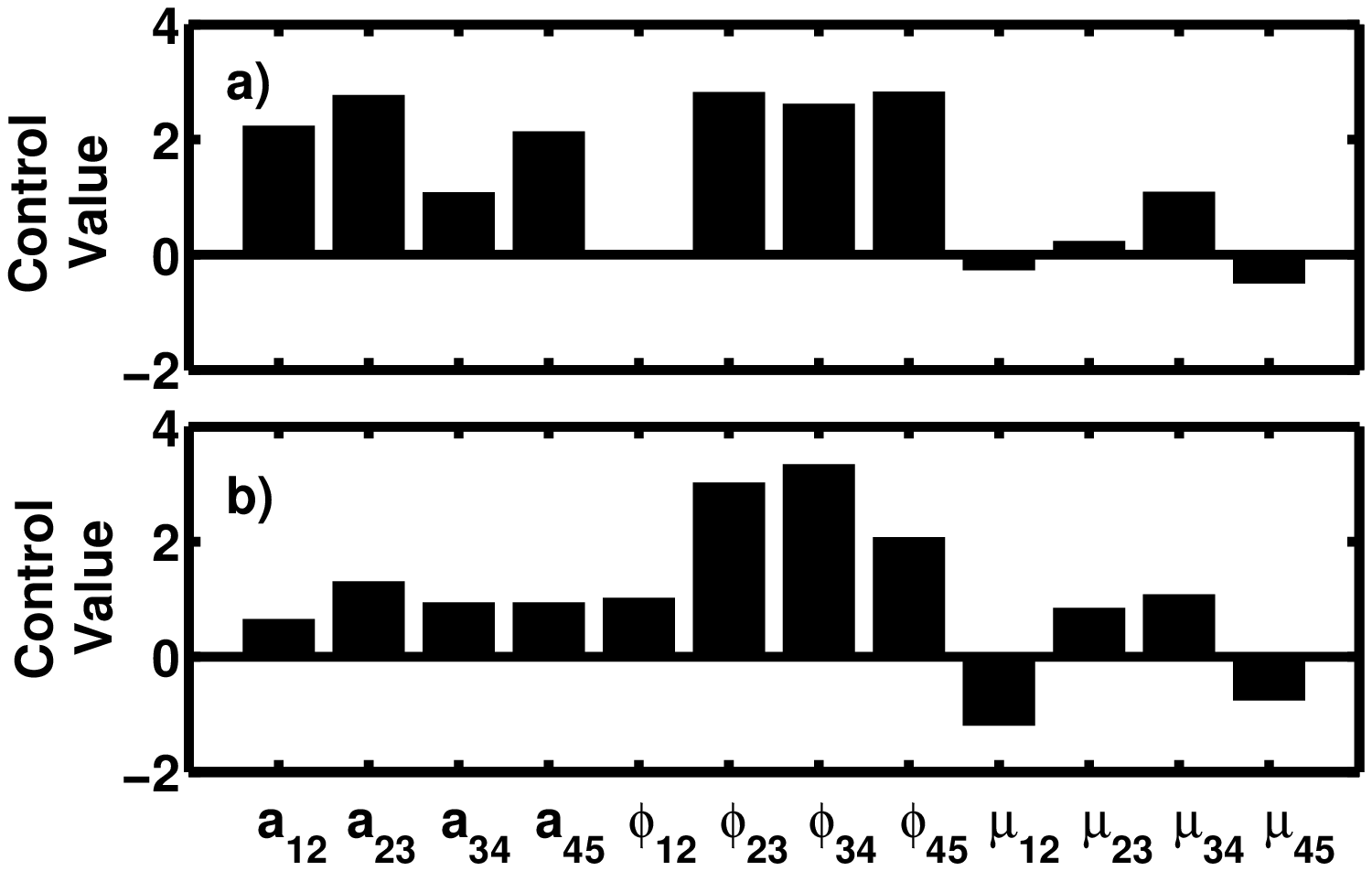}
\newline\newline Figure 3
\end{figure}

\clearpage

\begin{figure}
 \centering
 \includegraphics[scale=1.00]{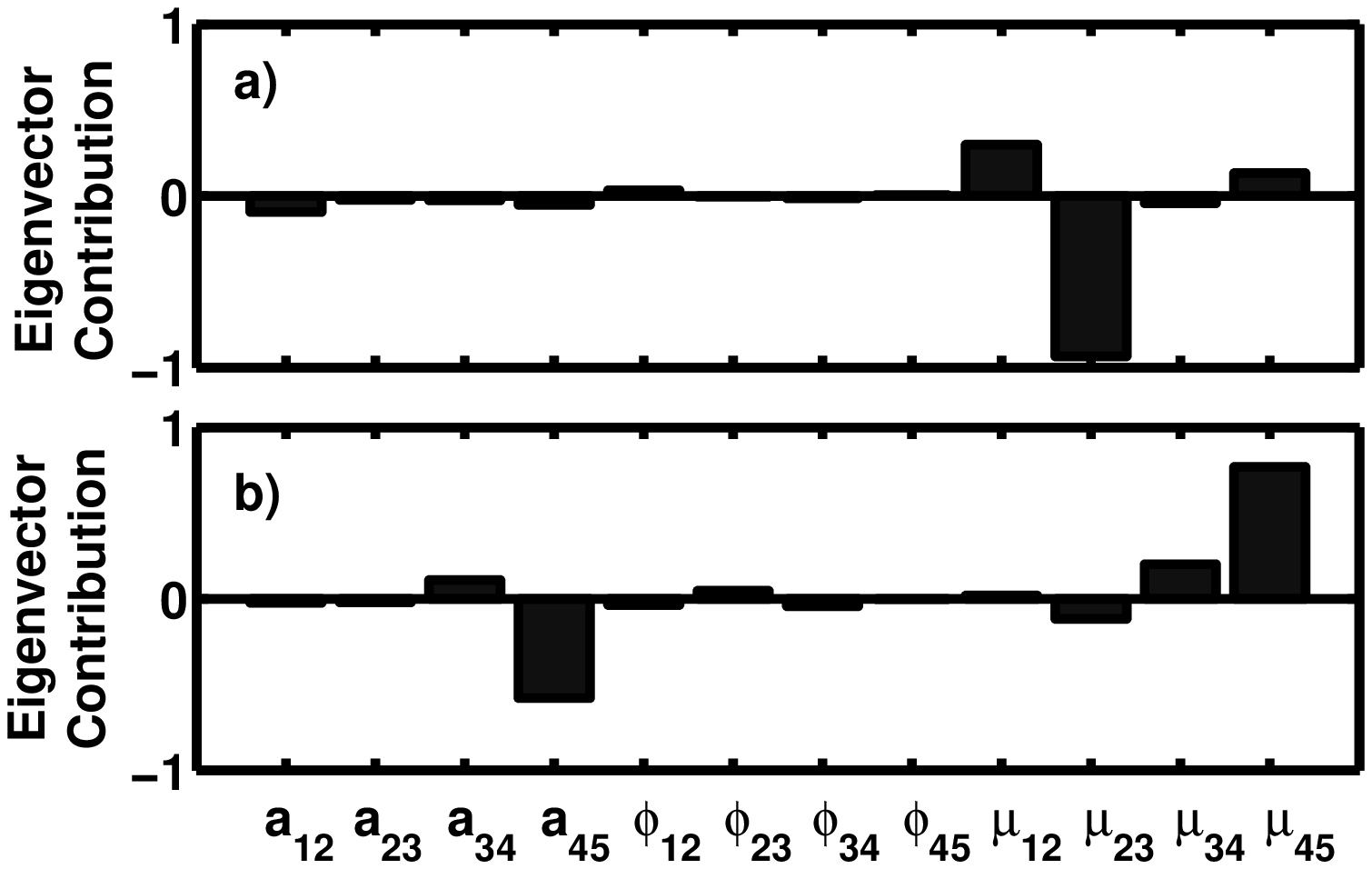}
\newline\newline Figure 4
\end{figure}

\clearpage

\begin{figure}
 \centering
 \includegraphics[scale=0.80]{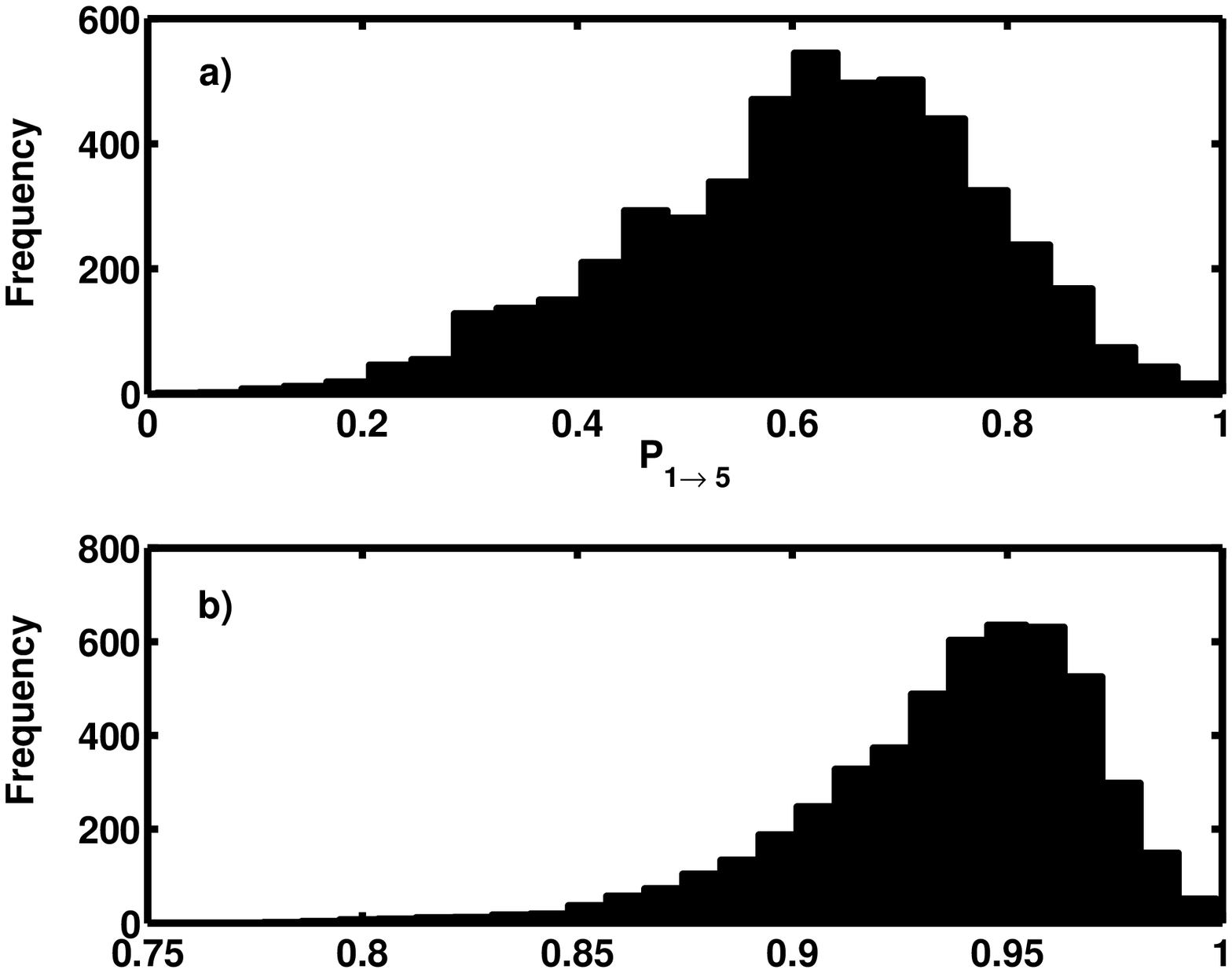}
\newline\newline Figure 5 
\end{figure}

\begin{figure}
 \centering
 \includegraphics[]{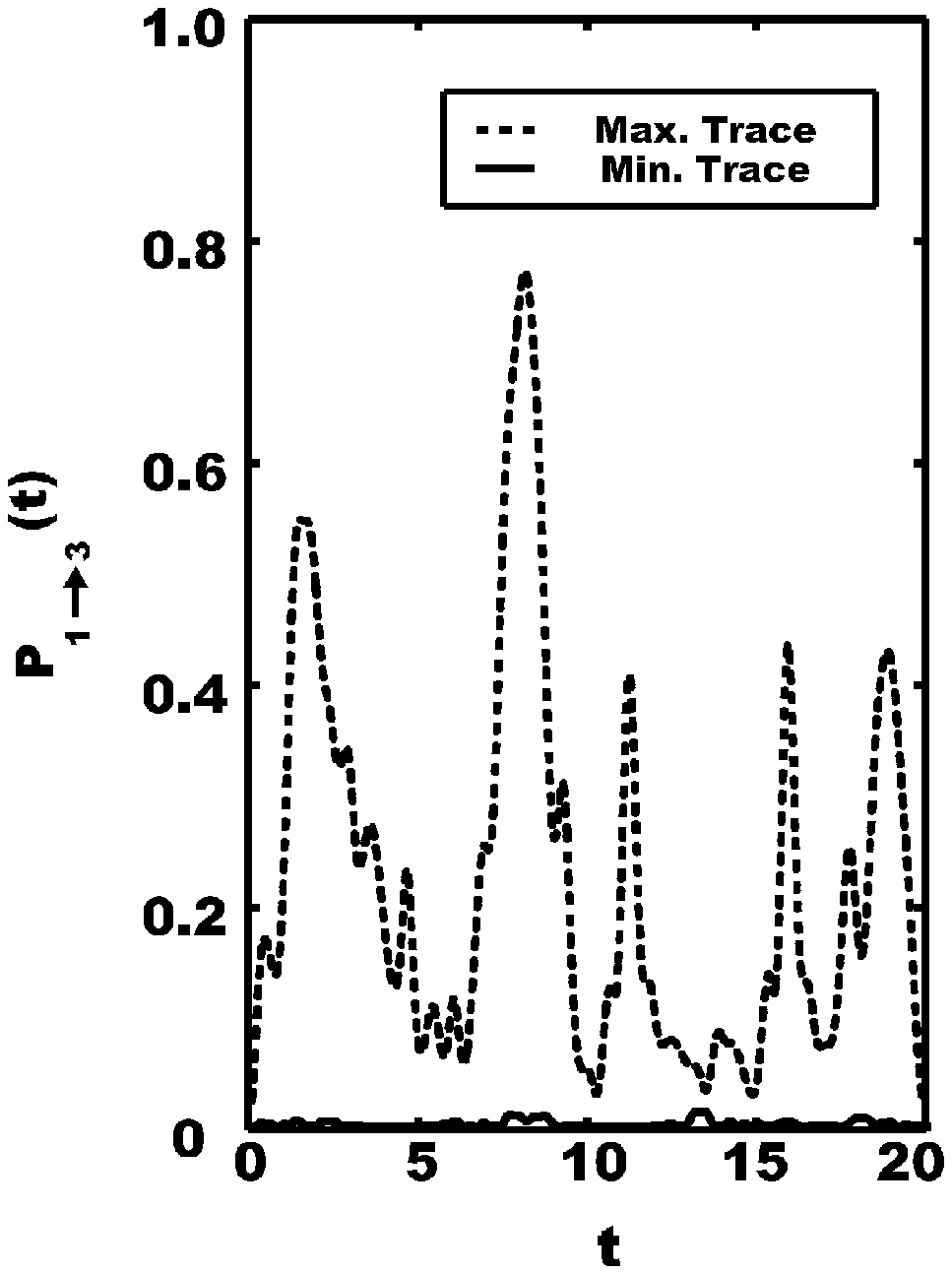}
\newline\newline Figure 6
\end{figure}
\clearpage

\begin{figure}
 \centering
 \includegraphics[scale=1.33]{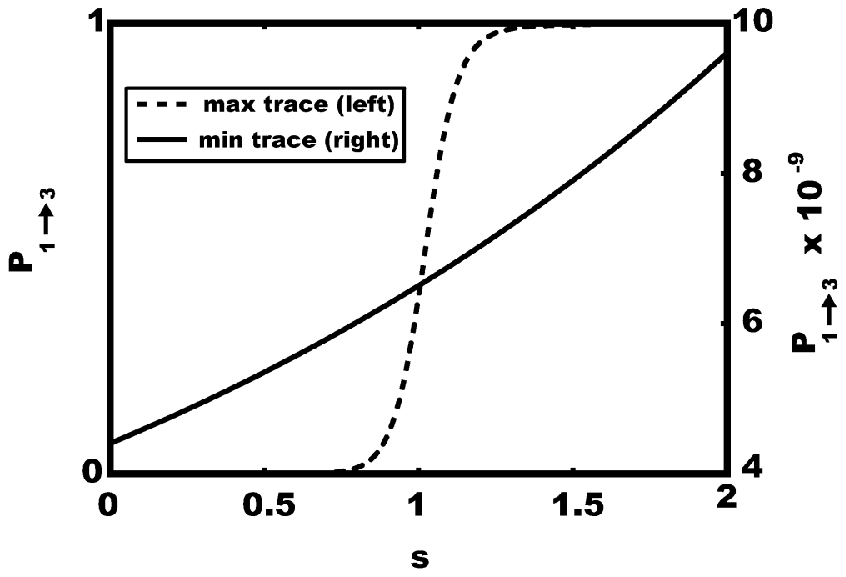}
\newline\newline Figure 7
\end{figure}

\clearpage
\noindent\textbf{Figure 1}\newline
\noindent 
Diagram of control variable options drawn from the Hamiltonian.   
(a) The traditional view of quantum control where the applied field is the
control and the Hamiltonian structure is fixed.
(b) The control is the Hamiltonian structure (e.g., energy level spacings and
couplings) and the corresponding 
dynamics are facilitated by a fixed applied control field.  In this case, the
landscape is explored by traversing the Hamiltonian structure controls.  
(c) Both the Hamiltonian structure and the applied field are the controls.  The
overall purview in (c) reveals the richness in considering controls through
access to molecular/material samples and external fields.  With a suitable
choice of controls, the entire observable landscape is subject to exploration,
including the level sets at the top and bottom.
\newline\newline

\noindent\textbf{Figure 2}\newline
\noindent The $P_{1\to2}=1$ level set at the top of the Hamiltonian structure
landscape for an $N=2$ state system in Section 4.1 with controls given by by
$E_{2}, \mu_{2}$ and $\mu_{12}$.  A fixed applied field $\epsilon(t)$ is present
facilitating the dynamics of the system.  Every point along each trajectory
corresponds to a control set $E_{2},\mu_{2},\mu_{12}$ that
accomplishes perfect yield under the influence of the same field.  
Each of the seven level set curves start with $E_{2}=0, \mu_{2} = 0$ and a
particular point along the $\mu_{12}$ axis.\newline\newline
\noindent\textbf{Figure 3}\newline
Values of the control variables for the nominal $\beta_{0}^{*}$ and robust
$\beta_{\textrm{rob}}^{*}$ control solutions for the example in Section 4.2 at
the top of the landscape for $\pp=1.0$.  These parameter settings correspond to
field amplitudes and phases, as well as the dipole couplings as discussed in the
text.  The control $\beta_{0}^{*}$ in (a) produces a Hessian trace of $-180.9$,
while the control $\beta_{\textrm{rob}}^{*}$ in (b) produces a Hessian trace of
$-30.3$.    \newline\newline
\noindent\textbf{Figure 4}\newline
Largest eigenvector of the Hessian at the top of the landscape ($\pp=1$) for the
nominal and robust control solutions (a) $\beta_{0}^{*}$ and (b)
$\beta_{\textrm{rob}}^{*}$, respectively, from the example in Section 4.2 and Fig. 3.  
In (a) the eigenvector for the nominal case has a corresponding eigenvalue of
$-69.7$ while 
in (b) the eigenvector for the most robust case has a corresponding eigenvalue
of $-14.1$. The nominal control solution shows sensitivity to nearly every
dipole value. The robust control solution shows a significant sensitivity 
to the amplitudes $a_{34}$ and $a_{45}$, as well as the corresponding dipole elements. \newline\newline
\noindent\textbf{Figure 5}\newline
Distribution of $P_{1\rightarrow 5}$ values after perturbing (a) the nominal
control solution $\beta_{0}^{*}$ and (b) the robust control solution
$\beta_{\textrm{rob}}^{*}$ with Gaussian distributed variations of each
control variable for the example discussed in Section 4.2 and Fig. 4.  The mean
$P_{1\rightarrow 5}$ value for (a) is $0.612$ with left standard deviation $0.174$.  
Note the distinct abscissa scales in (a) and (b). The mean
$P_{1\rightarrow 5}$ value for (b) is 0.936 with left standard deviation
$0.040$.  \newline\newline
\noindent\textbf{Figure 6}\newline
Population of state $|3\rangle$ over time, $P_{1\rightarrow 3}(t)$, for the
control field producing either a Hessian maximal or minimal trace at the bottom of the landscape
for the example in Section 4.3.  The minimal trace field induces dynamics that
keep the population in state $|3\rangle$ from ever exceeding $0.03$ while the
maximal trace field produces much more complex dynamics with large population
spikes over the time interval.  Both control fields result in a final value,
$P_{1\rightarrow 3} \approx 4\times 10^{-9}$ on the level set bottom at the
final time. \newline\newline 
\noindent\textbf{Figure 7}\newline
Results of two separate gradient ascents of the control landscape using Eq.
\ref{1st:climb} over the fixed range $s\in[0,2]$ starting near the bottom
($\pif < 10^{-8}$) of the applied field landscape for the example in Section 4.3
and Fig. 6.  The ascents start at different locations on
the landscape bottom corresponding to fields producing a Hessian trace of either maximal value 
(dashed: left-axis) or minimal value (solid: right-axis).  Both initial control fields produce $P_{1\rightarrow
3} < 10^{-8}$ at $s=0$ and, in both cases, the goal is to rapidly increase $P_{1\rightarrow 3}$. 
At $s=2$, the ascent originating from the maximal trace
field (trace = $8\times 10^{3}$) reaches a value of $P_{1\rightarrow 3} \approx
0.999$, whereas the ascent originating from the minimal trace field (trace =
$0.31$) just reaches a value of $P_{1\rightarrow 3}\approx 9.5\times 10^{-9}$. 
On the scale of the abscissa, the ascent for the minimal trace field reaches
$P_{1\rightarrow 3} > 0.9$ only at $s\approx 68$.\newline\newline
\clearpage

\end{document}